\documentclass{elsart}
\usepackage{latexsym}

\def\kms {\hbox{km{\hskip0.1em}s$^{-1}$}} % km/s
\def\ee #1 {\times 10^{#1}}          % \ee p       10^p
\def\ut #1 #2 { \, \mathrm{#1}^{#2}} % \ut unit p  unit^p
\def\u #1 { \, \mathrm{#1}}          % \u unit     unit

\def\kms{km s$^{-1}$}

\def\psec           {$.\negthinspace^{s}$}

\def\pdeg           {$.\kern-.25em ^{^\circ}$}
\def\degree{\ifmmode{^\circ} \else{$^\circ$}\fi}
\def\ee #1 {\times 10^{#1}}          % \ee p       10^p
\def\ut #1 #2 { \, \textrm{#1}^{#2}} % \ut unit p  unit^p
\def\u #1 { \, \textrm{#1}}          % \u unit     unit

\def\dsec   {\hbox{$.\!\!^{\rm s}$}}            % Second over dot
\usepackage{natbib}

\usepackage{graphicx}
\usepackage{amssymb}

\begin{document}

\begin{frontmatter}



\title{The Nature of Nonthermal X-ray  Filaments
Near the Galactic Center}


\author{F. Yusef-Zadeh}
\address{Dept. Physics and Astronomy, Northwestern University, 2145 Sheridan
Road, Evanston IL. 60208 }

\author{M. Wardle}
\address{Department of Physics, Macquarie University, Sydney, NSW 2109,
Australia}

\author{M. Muno}
\address{Dept.  Physics and Astronomy, UCLA, Los Angeles, CA 90095}

\author{C. Law}
\address{Northwestern University, 2145 Sheridan Road, Evanston IL. 60208 }

\author{M. Pound}
\address{Dept of Astronomy, University of Maryland, MD 20742}

\begin{abstract}
% Text of abstract

Recent Chandra and XMM-{\it Newton} observations reported evidence of two 
X-ray filaments G359.88-0.08 (SgrA-E) and G359.54+0.18 (the ripple 
filament) near the Galactic center.  The X-ray emission from these 
filaments has a nonthermal spectrum and coincides with synchrotron 
emitting radio sources.  Here, we report the detection of a new X-ray 
feature coincident with a radio filament G359.90-0.06 (SgrA-F) and show 
more detailed VLA, Chandra and BIMA observations of the radio and X-ray 
filaments.  In particular, we show that radio emission from the nonthermal 
filaments G359.90-0.06 (SgrA-F) and G359.54+0.18 (the ripple) has a steep 
spectrum whereas G359.88-0.08 (SgrA-E) has a flat spectrum. 
The X-ray emission from both these sources could be due to synchrotron 
radiation. However, 
given
that the 20 \kms molecular cloud,  with its intense 1.2mm dust emission,
lies in the vicinity of SgrA-F, it is possible that the X-rays could
be produced by inverse Compton scattering of far-infrared photons from
dust by the relativistic electrons responsible for the radio
synchrotron emission.
The production of X-ray emission from ICS allows an estimate of the 
magnetic field strength of ~0.08 mG within the nonthermal filament.  This 
should be 
an important parameter for any models of the Galactic center nonthermal 
filaments.

\end{abstract}


\end{frontmatter}

% main text
\section{Background}
\label{}

One of the characteristics that defines the center of our Galaxy is a
significant amount of thermal and nonthermal material distributed in
this region.  The nonthermal emission arises mainly from a variety of
synchrotron sources such as SNRs, unusual filamentary structures and a
diffuse 1$^0$-scale lobe running perpendicular to the Galactic plane
(e.g., Law et al.  2004).  On the other hand, thermal emission arises
from stellar sources, HII regions, and the dense clouds of dust and
gas pervading the inner two degrees of the Galactic center (see Morris
and Serabyn and references therein 1996).  One of the questions that
naturally arises from these observations is whether the inverse
Compton scattering (ICS) of infrared radiation by relativistic
particles is an important contribution to the nonthermal X-ray
emission from this region of the Galaxy.  Here we investigate this
possibility on a local scale and examine the application of ICS in
three nonthermal filaments that are detected in both radio and X-ray
wavelengths.  All the filaments lie in the vicinity of molecular
clouds characterized by their high density $\approx 10^4-10^5 \ut cm
-3 $ and high temperature $\approx$ 70\,K (e.g., Morris and Serabyn
1996; Martin et al.  2004).

\section{Radio Continuum, Molecular Line and X-ray Observations}
\label{}

A nonthermal X-ray source XMM J174540--2904.5 was first reported by
Sakano et al./ (2003), based on 50 ks exposure {\it Chandra}
 and XMM-{\it Newton} observations of the Galactic center.
XMM J174540--2904.5 has a hard spectrum with an extent of
$\approx15''$ and is displaced by $\approx4'$ south of the well-known
compact radio continuum source SgrA$^*$ at the dynamical center of the
Galaxy.  XMM J174540--2904.5 coincides with a nonthermal radio
feature, the SgrA-E ``wisp" G359.88-0.08 (Ho et al.\ 1985; Yusef-Zadeh
and Morris 1987).  Sakano et al.\ (2003) identify XMM J174540--2904.5
as a nonthermal radio filament with characteristics similar to those
observed throughout the Galactic center (e.g., Yusef-Zadeh, Morris and
Chance 1984; Morris and Serabyn 1996).  Another interpretation was
recently given by Lu, Wang and Lang (2003) who suggest XMM
J174540--2904.5 traces a nonthermal pulsar wind nebula.

We carried out a detailed radio and X-ray study of the SgrA-E source
as well as a new source, SgrA-F, using the long exposure 600ks Chandra
observation of the Galactic center (Baganoff et al.\ 2003; Muno et
al.\ 2003).  Radio continuum (projects AH-1, AA31 and AY131) were carried 
out
at 20cm, 3.6 and 2cm using the Very Large Array (VLA) of the National
Radio Astronomy Observatory.\footnote{The National Radio Astronomy
Observatory is a facility of the National Science Foundation, operated
under a cooperative agreement by Associated Universities, Inc.} SgrA-E
and F sources were also imaged in the CS(2-1) transition at 97.980968
GHz with the Berkeley-Illinois-Maryland Association (BIMA)
interferometer\footnote{The BIMA interferometer is operated under a
joint agreement between the University of California, Berkeley, the
University of Illinois, and the University of Maryland with support
from the National Science Foundation.} during the 2002-2003 observing
season.  The sources were observed with two antenna pointings centered
at $(RA, Dec)_{2000.0} =$ \mbox{(17$^h$45$^m$40\psec 4,
$-$29$\degree$04$^\prime$30)} and \mbox{(17$^h$45$^m$38\psec 3,
$-$29$\degree$03$^\prime$30)} with an LSR velocity of 20.0 \kms\/.  We
observed in the C configuration of the array, sampling spatial
frequencies from 2 k$\lambda$ to 29 k$\lambda$.  Absolute flux
calibration was derived from observations of Mars immediately before
or after the source track.  The quasar 1733-130 was used as the phase
calibrator and secondary flux calibrator.  We also obtained
fully-sampled single-dish maps covering the region imaged with the
array in CS(2-1) using the Five Colleges Radio Astronomy Observatory 
(FCRAO) 
14~m telescope, which has a 45$''$ (FWHM) beam at 115 GHz.  The CS
single-dish data were resampled to the BIMA velocity resolution and
put on the same flux scale as the BIMA observations using a conversion
from antenna temperature of 43 Jy/K (M. Heyer 1999, private
communication).  The single-dish data  were then 
linearly combined with
the ``dirty'' interferometric maps and jointly deconvolved
(Stanimirovic et al.  1999).  The $1 \sigma$ rms noise per channel in
the final combined map is 0.18 Jy/beam.  (1 Jy/beam = 1.1 K at the CS
frequency and synthesized beam size).

Twelve separate pointings toward the Galactic center have been carried
out using the Advanced CCD Imaging Spectrometer imaging array (ACIS-I)
aboard the {\it Chandra X-ray Observatory}, in order to monitor Sgr
A$^*$ (Baganoff et al.  2003) The ACIS-I is a set of four 1024-by-1024
pixel CCDs, covering a field of view of 17$'\times17'$.  The CCDs
measure the energies of incident photons, with a resolution of 50--300
eV (depending on photon energy and distance from the read-out node)
within a calibrated energy band of 0.5--8~keV. The angular resolution
at the aim-point of the telescope is better than 0.5$''$ at 6.4~keV,
while within the inner 4$'$ from the aim-point it is always better
than 2$''$.  We obtained spectra of SgrA-E and F using the techniques
described in detail in Muno et al.  (2003).  In brief, we created an
event list for each observation in which we corrected the pulse
heights of each event for the position-dependent charge-transfer
inefficiency.  The final total live time was 626 ks.  We then applied
a correction to the absolute astrometry of each pointing using three
Tycho sources detected strongly in each {\it {Chandra}} observation
(Baganoff et al.  2003), and re-projected the sky coordinates of each
event to the tangent plane at the radio position of Sgr A$^*$ in order
to produce a single composite image.  The image was searched for point
sources using the program {\it{wavdetect}} in three energy bands
(0.5--8~keV, 0.5--1.5~keV, and 4--8~keV) using a significance
threshold of $10^{-7}$.  SgrA-E and F were both detected as part of
this search, the results of which are listed in the electronic version
of the catalog (Muno et al.  2003).

\section{G359.88-0.08 (SgrA-E) and G359.90-0.06 (SgrA-F)}
\label{}

Figure 1 displays contours of X-ray emission from XMM J174540--2904.5
superimposed on a grayscale continuum image at 2cm with a resolution
2.2$''\times1.3''$.  The coincidence of X-ray and radio emission from
the elongated source SgrA-E is clearly noted.  The bright circular
feature SgrA-G to the southwest corner is an HII region (Ho et al.
1985; Yusef-Zadeh and Morris 1987).  The radio emission from SgrA-E
extends for $\sim2'\times15''$ with a peak brightness of 5.5 mJy
beam$^{-1}$ at the apex of this ``banana-like'' structure.

We also detect an additional X-ray source coincident with SgrA-F
located to the northwest of SgrA-E in Figure 1.  Figure 2 shows a
close-up view of the contours of total intensity at 20cm superimposed
on an X-ray grayscale image of SgrA-F source.  The radio image of
SgrA-F consists of two parallel subfilaments, one of which is seen to
have an X-ray counterpart with a size of $\approx6" \times 3''$
(length $\times$ width).  The coincidence between radio and X-ray
emission from the fainter radio subfilament of SgrA-F is detected at
$(RA, Dec)_{2000.0} = 17^{\rm h} 45^{\rm m} 38^s.09, -29^{\circ} 03'
20.35"$.  The lack of X-ray emission from the southern peak of SgrA-F
in Figure 2 is in part due to the low-sensitivity region experienced
in the chip gaps of the ACIS observations.  Most of the X-ray sources
have a shape similar to that expected from the Chandra point spread
function (PSF), whereas SgrA-F is clearly more extended than the PSF.
Therefore, it seems unlikely that it is a chance association with an
extended radio source.  Figure 3 shows high resolution radio and X-ray
images of SgrA-E. The X-ray and radio emission from SgrA-E are
well-correlated, both displaying a sharp eastern edge and a diffuse
western edge.  The radio filament SgrA-E is much more elongated in its
extent than its X-ray counterpart, these images show clear
morphological evidence that the X-ray and radio emission from the
brightest segment of SgrA-E arise from the same region.  Similarly,
although with limited X-ray data affected by the chip gap, the
extended X-ray emission appears to be correlated with the fainter
component of the filamentary structure SgrA-F in radio wavelengths.
  
Figure 4 shows the X-ray spectra of SgrA-E and F. Both spectra are 
featureless, and emit copious X-ray emission between 4 and 8 keV. The 
X-ray spectra were modeled in {\tt XSPEC}, using a power-law continuum 
absorbed at low energies by gas (the model {\tt phabs} in {\tt XSPEC}) and 
dust (the model {\tt dust}) in the interstellar medium (Baganoff 2003).  
The column depth of dust was set to $\tau = 0.485 \cdot N_{\rm H}/(10^{22} 
{\rm cm}^{-2})$, and the halo size to 10 times the PSF size.  The spectrum 
of SgrA-E contained 3600 net counts, and is consistent with a photon index 
$\Gamma=2\pm0.5$ (where $\rm I_{\rm E}\propto E^{-\alpha}$ and 
$\alpha=\Gamma-1)$ power-law absorbed by $N_{\rm H} = (2.5 \pm 0.4) \times 
10^{23}$ cm$^{-2}$ (uncertainties are 90\% confidence for a single 
parameter of interest).  The total observed and unabsorbed 2--8~keV flux 
from SgrA-E in the 380 $\Box''$ extraction region is $2.5 \times 10^{-13}$ 
and $1.3 \times 10^{-12}$~erg cm$^{-2}$ s$^{-1}$, respectively.  This 
corresponds to X-ray luminosity L$_X\approx10^{34}$ ergs s$^{-1}$ at the 
distance of 8 kpc.  Assuming a bremsstrahlung continuum instead yields a 
similar absorption column, the best-fit value on the temperature of the 
emitting egion is $kT = 14$~keV. with a 90\% lower limit $ 8$~keV and no 
upper limit.

The X-ray spectrum of SgrA-F is poorly constrained, because it is
based on only 167 net counts.  The 90\% confidence intervals on the
spectral parameters were $\Gamma$ = (0.8, 8.2) and $N_{\rm H} = (1.7,
7.9)\times 10^{23}$ cm$^{-2}$.  If we assume an absorption column of
$2.5 \times 10^{23}$ cm$^{-2}$, the photon index is still only
constrained to $\Gamma$ = 2.0 (1.1, 3.0).  The total observed and
unabsorbed 2--8~keV flux in the 30 $\Box''$ extraction region is $9.4
\times 10^{-15}$ and 5.7 $\times 10^{-14}$~erg cm$^{-2}$ s$^{-1}$,
respectively.  The uncertainty on the spectral index of SgrA-F is
large, but the range of values is consistent with that seen from
SgrA-E.

The spectral index distribution measured at radio wavelengths shows
that SgrA-E has a steeper spectrum than  SgrA-F. In particular,
$\alpha\approx-0.17\pm0.1$ is based on 2 and 6cm images for SgrA-E
whereas $\approx -0.80\pm0.1$ is measured between 6 and 3.6 cm for
SgrA-F. The spectral index distribution between 6 and 2cm was
constructed by using an identical {\it {uv}} range between 0.4 and 40
k$\lambda$.

The flat spectrum of Sgr A-E suggests an interaction with a molecular
cloud.  Indeed, an interaction between the 20 \kms molecular cloud and SgrA-E
and F has been suggested earlier in an NH$_3$ line study of this
region (Coil and Ho 2000).  To examine the interaction hypothesis,
high-resolution CS (2--1) observations of this region were carried out
with BIMA with a spatial resolution of $20''\times5.8''$.  Figure 5
shows contours of integrated line emission between 0 and 35 \kms
superimposed on a grayscale 6cm continuum image (shown in black). 
In order to highlight the small-scale molecular features, in this
 figure we show the BIMA map without the FCRAO zero-spacing
 data. However, Figures 6a,b are  derived from the combined data. The CS 
(2--1)
emission appears clumpy on a scale of few arcseconds and its
distribution is concentrated in the vicinity of SgrA-E and F. The most
striking result is the recognition of a molecular filament which
appears to be the counterpart to the northern extension of the
banana-shaped structure of SgrA-E near$(RA, Dec)_{2000.0} = 17^{\rm h}
45^{\rm m} 39\dsec5, -29^\circ 03^\prime 45^{\prime\prime}$.  A slice
cut perpendicular to the long axis of the molecular feature enveloping
SgrA-E, as displayed in Figure 6a, shows that the velocities range
between 10 and 40 \kms.  This position-velocity diagram indicates a
velocity shift of about 5 \kms across the width of this molecular
filament with an extent of about 15$''$.  This is also the region
where NH$_3$ line emission from the 20 \kms molecular cloud has been
detected with H$_2$ column density of $\approx10^{24}$ cm$^{-2}$ (Coil
and Ho 2000).  The CS (2--1) emission probes high density gas
consistent with the high column density noted there.  The
brightest CS (2--1) emission, as presented by a cross in Figure 5, is
adjacent to the region where X-ray emission is detected from SgrA-E. A
velocity line profile of the peak position, as shown in Figure 6b,
indicates a total linewidth of about 40 \kms from a size of about
3$''\times3''$.

Additional support for the proximity of the 20 \kms GMC (G359.98-0.07
or M-0.13-0.08) to SgrA-E and F comes from the distribution of dust
emission.  Figure 7 shows contours of the brightest 1.2mm dust
emission from the southern segment of the 20 \kms cloud (Zylka et al.
1998) superimposed on a grayscale 3.6cm continuum emission from SgrA-E
and F. Altogether, Figures 5 to 7 indicate that the bulk of CS (2--1)
emission with a velocity of $\approx$20 \kms is associated with the 20
\kms GMC. Moreover, there is compelling kinematic and morphological
evidence from CS (2--1) and 1.2mm dust emission from the 20 \kms GMC
that suggests physical interaction between the 20 \kms
cloud and SgrA-E and F.

\section{G359.54+0.18 (the ripple filament)}

The ripple filament is another isolated filamentary structure in the
Galactic center region with a terminus that flares toward a 
HII complex.  Linear polarization measurements show that the local
magnetic field traces the filament (Yusef-Zadeh, Wardle and Parastaran
1997).  A dense molecular cloud and an HII region are observed at one
end of the filament suggesting that bending of the filament or a
"ripple" is produced as a result of interaction (Bally and Yusef-Zadeh
1989; Staguhn et al.  1998).  A recent 20cm radio continuum image of
G359.54+0.18 shows that the length of this filament extends for about
15$'$.  However, the brightest portion of the filament which is
resolved into two parallel subfilaments extends only for 5$'$
(Yusef-Zadeh,  Hewitt and Cotton 2004).

The detection of X-ray emission from the brightest portion of the
ripple filament G359.54+0.18 was first reported by Lu, Wang and Lang
(2003) as part of the Galactic center survey carried out by Chandra.
The ripple filament was observed nearly on-axis by Chandra (ObsID
2268) for 11-ksec.  We calibrated this observation with the standard
pipeline processing and analyzed with CIAO 2.2.1.  A background
spectrum was extracted with the same region as the source spectrum
from a customized "blank-sky" file \footnote{see
http://cxc.harvard.edu/ciao/threads/acisbackground}.  X-ray emission
is seen from the region where the radio filament breaks up into two
subfilaments.  Figure 8a shows the brightest portion of G359.54+0.18
at radio and X-rays.  X-ray emission from G359.54+0.18 is superimposed
on a grayscale smooth Chandra image in the top panel; contours of
X-ray emission superimposed on a grayscale 6cm image of the northern
and southern subfilaments are shown in the bottom panel.  A weak X-ray
filament with an extent of 1$'$ is noted exactly along the northern
subfilament.  The brightness temperature of the two subfilaments at
6cm are similar but, interestingly, only the northern subfilament
appears to have an X-ray counterpart.  Figure 8b shows the weak X-ray
spectrum, and the best-fit absorbed powerlaw model.  The 60 count
spectrum is extracted from an 11-ksec observation and is fitted with
an absorbed powerlaw.  The best-fit values (with 90\% errors) are
$N_H=11.4^{+14}\times10^{22}$ cm$^{-2}$ and
$\Gamma=4.1^{+4.4}_{-2.7}$, which gives an unabsorbed 0.5 to 8 keV
flux of $3\times{10^{-12}}$ ergs
cm$^{-2}$ s$^{-1}$ with upper and lower limits of 
$10^{-8}$ and 3$\times10^{-13}$ ergs 
cm$^{-2}$ s$^{-1}$ , respectively.  Note that the 
unabsorbed 
flux decreases with
lower values of $\Gamma$ (i.e., a flatter spectrum).  The radio flux
density from the region of X-ray emission is 60 mJy at 6cm
corresponding to $\nu F_\nu \approx 2\times10^{-14}$ erg s$^{-1}$
cm$^{-2}$ which is about two orders of magnitude weaker than its X-ray
counterpart.

We measured spectral indices along the ripple filaments of
$\alpha\approx=-0.8$ between 6 and 20cm and $\approx-2$ between 6 and
3.6cm.  The spectral index has previously been determined to be
$\alpha=-0.8$ between 90 and 20cm (Anantharamaiah et al.\ 1991).
Figure 9a,b show the spectral index distribution of an identical
region of the ripple filament between 20 \& 6cm and 6 \&3.6 cm,
respectively.  No noticeable difference is noted in the spectral index
values of the northern and southern subfilaments.  The break in the
spectral index is similar to that of the "northern thread" filament
G0.08+0.15 which has $\alpha\sim-0.5$ at low frequencies but
steepening to $\alpha=-2$ between 6 and 2cm wavelengths (Lang, Morris
\& Echevarria 1999).

\section{Discussion}
\subsection{The SgrA-E Filament: Synchrotron Mechanism}

Sakano et al (2003) suggest that the X-ray emission from XMM
J174540--2904 is due to synchrotron radiation emitted from TeV
electrons.  In particular, they find that the X-ray spectrum from
SgrA-E is consistent with steepening by 0.5 in slope between radio and
X-ray wavelengths.  Our improved spectral index determinations at
radio and X-ray energies differs from earlier X-ray and radio spectral
measurements (Sakano et al.  2003; Ho et al.  1985).  On the X-ray
side, Sakano et al.  (2003) used older absorption cross sections and
excluded the dust contribution which increases their derived
absorption column by a factor of 2.  Our derived photon index value is
more accurately measured than earlier work by Sakano et al.  (2003)
who found $\Gamma=2 (1.1, 3.1)$.  However, their best-fit power law
with their uncertainties are consistent with the new fitted values,
$\Gamma=2\pm0.5$.  Since our new Chandra observations is about 20
times more sensitive than previous observations, uncertainties of the
spectral fit are smaller and should be more reliable.  On the radio
side, we presented new high resolution radio continuum data and
measured flux densities from the region with an X-ray counterpart at a
number of wavelengths between 2 and 20cm.  Our radio spectral index is
$-0.17$, flatter than the value of $-0.4$ obtained by Sakano et al.\
(2003) using data sets with different resolutions and {\it uv}
coverage.

Our radio and X-ray data show a steepening in the spectral index of
$\sim$ 0.8 from radio to X-rays.  Such a change in the spectral index
value is consistent with the synchrotron picture suggested by Sakano
et al.  (2003).  In this picture, the elongated nature of SgrA-E which
extends for more than 2$'$ compared to 30$''$ extent of the X-ray
filament suggests that synchrotron cooling losses are responsible for
the origin of the different morphology between radio and X-rays (see
Figure 3).  Assuming a magnetic field of 100$\mu$G and electron
energies $\approx$TeV, X-ray emission has a lifetime of $\approx10^3$
years which is much shorter than the lifetime of GeV electrons
producing radio emission (e.g., Beck and Krause 2005).  During this
time, the diffusion of X-ray emitting particles with a velocity of 300
\kms can explain the 30$''$ length of the filament which corresponds
to about $\sim$1 pc at the 8kpc distance of the Galactic center.

What could be the origin of the acceleration of particles to
relativistic energies?  The flat spectrum of the region traced by
SgrA-E at radio frequencies provides an insight as what the cause
might be.  This flat spectrum is reminiscent of the W28 SNR which is
known to be interacting with an adjacent molecular cloud (Dubner et
al.  2002).  The flat spectral index distribution is consistent with
the Fermi acceleration mechanism predicting a flatter spectrum when a
shock wave drives through a molecular cloud (Bykov et al.  2000).
This picture is supported by the distribution and kinematics of the CS
(2-1) molecular gas associated with the 20 \kms cloud in the vicinity
of SgrA-E.

Also, we note a number of stellar X-ray sources detected in the region
near SgrA-E (Muno et al.  2003).  Based on the low exposure X-ray
image of G359.88-0.08, Lu, Wang and Lang (2003) have suggested that a
compact X-ray source CXOU J174539.6-290413 is located to the northern
tip of XMM J174540--2904.5 and that it may coincide with a pulsar
responsible for producing X-ray and radio emission from SgrA-E.
However, the deeper observations presented here (see Figure 3a) do not
confirm the presence of a point-like X-ray source at the tip of the
structure.

\subsection{The SgrA-F filament: synchrotron or inverse Compton emission}

Unlike the SgrA-E filament, which has flat and steep spectra in radio
and X-rays, respectively, the SgrA-F filament has a steep spectrum at
both radio ($\alpha \approx 0.8$) and X-ray wavelengths
($\alpha=\Gamma-1\sim 1$).  In principle the radio and X-ray spectra
could be segments of a single synchrotron spectrum with $\alpha
\approx 0.87$ between radio and X-rays.  However, as the radio
spectrum of Sgr A-F is much steeper than that of Sgr A-E, and given
that the 20 \kms molecular cloud,  with its intense 1.2mm dust emission, 
lies in the vicinity of SgrA-F, it is possible that the X-rays could
be produced by inverse Compton scattering of far-infrared photons from
dust by the relativistic electrons responsible for the radio
synchrotron emission.

To estimate the IC emission we adopt a greybody FIR spectrum from
warm dust at temperature $T_d$ of the form
$$
I_\nu = B_\nu(T_d) \times (1-\exp(-\tau_\nu))
$$
where $B_\nu$ is the Planck function and $\tau_\nu = \tau_0
(h\nu/kT_d)^\beta$ is the dust optical depth.  Then the X-ray flux at 1\,keV
produced by inverse Compton scattering off a synchrotron source of
flux $S_\nu \propto \nu^{-\alpha}$ is
$$
F_X = 7.18\ee -19 \; \Gamma'(\alpha+\beta+3)\, \tau_0
      \frac{T_{50}^{3+\alpha}}{B_\mathrm{mG}^{1+\alpha}} S_\mathrm{mJy}
       \quad\mathrm{erg\,cm^{-2}\,s^{-1}\,keV^{-1}}
$$
where $\Gamma'$ is the Gamma function, $S_\mathrm{mJy}$ is the
synchrotron flux at 731\,MHz, $T_{50} = T_d / 50$\,K, and
$B_\mathrm{mG}$ is the magnetic field strength in milligauss.

The thermal dust emission in the vicinity of SgrA-F is estimated from
observations of the 20 \kms\ cloud at 1.2mm with IRAM (Zylka et al.\
1998), and 850 and 450 $\mu$m with SCUBA (Pierce-Price et al.\ 2003).
These measurements are on the Rayleigh-Jeans tail of the Planck
function and so poorly constrain $T_d$ to be in the range $\sim
30-50$\,K, but characterize the frequency dependence of $\tau_\nu $,
which must scale as $\nu^2$ (i.e., $\beta=2$).  The fluxes in a 36$"$
radius circle centered on l=359$^0$.891 , b= -0.0731$^0$ gives
$\tau_0 \approx 0.065\, T_{50}$.

For a measured synchrotron flux of 25 mJy from SgrA-F at 2cm with a
$\nu^{-0.75}$ spectrum, $S_\mathrm{mJy} = 240$.  For a magnetic field
strength 1mG, the estimated ICS X-ray fluxes at 1 keV are
$8.8\times10^{-16}$ or $8.0\times10^{-17}$ erg cm$^{-2}$ s$^{-1}$
keV$^{-1}$ for $T_d = 50$\,K, and $30$\,K respectively.  These nominal
values are less than the observed flux of $3.2\times10^{-14}$ erg
cm$^{-2} s^{-1} keV^{-1}$ from the SgrA-F filament, but scale as
B$^{-1.75}$, and  match if $B\approx0.13$\,mG (T$_d=50$\,K) or 0.033 mG
(T$_d=30$\,K).  A recent ISO survey of the Galactic center molecular
clouds suggest the presence of warm and a cold dust
(Rodriguez-Fernandez et al.  2004).  Most of the clouds show dust
temperatures less than 40\,K. This suggests that the relatively large
X-ray flux from Sgr A-F would require a low magnetic field of
0.03-0.13 mG.

The equipartition magnetic field depends on the synchrotron flux, the
low-energy cutoff of the relativistic electron energy spectrum and the
depth of the source.  Assuming the depth is equivalent to 10$''$, the
equipartition field is 0.20mG for a cutoff of 10 MeV or 0.14mG for a
cutoff of 100 MeV. Note that this assumes that no relativistic protons
are present.  If the depth is 100$''$ which is higher than the
transverse dimension of the filament, the equipartition fields are
0.11mG and 0.07mG for cutoffs of 10 and 100 MeV respectively.

To summarize, the X-ray flux can be matched by inverse Compton
emission for a dust temperature of 50K and field strength 0.08mG. This
field might be in energy equipartition with the electrons provided
that the depth of the source is equivalent to 100$''$ (4pc at the
distance of 8kpc) and there are no relativistic protons present.
However, almost all the Galactic center radio filaments have
transverse dimensions which are typically $10''$ including the width
of SgrA-F. Thus, the estimated magnetic field is sub-equipartition.

\subsection{The Ripple filament}

The kinematical and morphological arguments for the interaction of the
ripple filament G359.54+0.18 with molecular gas and HII region have
been given by Staguhn et al.\ (1998).  
The break in the radio spectral index of the ripple from
$\alpha\sim-0.8$ to $-2$ implies that synchrotron emission from a
single population of relativistic electrons cannot be responsible since
the required radio to x-ray index is $\sim -0.85$.
Using the break
frequency of 5 GHz corresponding to 6cm and a magnetic field of 0.07
mG, the synchrotron lifetime of the ripple filament is 8$\times10^5$
years.

Inverse Compton scattering of the FIR radiation emitted by the dust
grains in the molecular clouds associated with the ripple filament
does not appear to be significant: the flux predicted in this model is
only comparable to the unabsorbed 0.5--10 keV flux, $\sim 3\times
10^{-12} \u erg \ut cm -2 \ut s -1 $, and even then only with the
marginal assumptions that the magnetic field is 0.07 mG and dust
temperature is 50K. This dust temperature is similar to the hot
component of dust temperature measured with ISO (Rodriguez-Fernandez
et al.  2004).  In addition, the X-ray emission arises only from the
northern subfilament (see Figure 8) raising the question of why in an
ICS the southern subfilament with the same radio brightness
temperature and similar FIR radiation field does not have an X-ray
counterpart?  One explanation is variations of the magnetic field and
particle density between the northern and southern subfilaments.  A
more detailed sensitive measurements should be important in testing
this hypothesis.

We also examined the possibility that the X-ray emission may arise
from synchrotron self-Compton scattering (SSC).  However, the
synchrotron emission from the filaments in radio wavelengths is too
weak to be important in the SSC model for the following reasons: The
ratio of SSC emission to inverse Compton scattering of a diluted
Black Body intensity W$\times B_{\nu}(T)$ is 2.5$\times10^{-4}$
assuming that the temperature of a molecular cloud is 10K. The
brightness temperature of G359.54+0.18 at 6cm is 50K and the
synchrotron radiation spectrum depends on $\nu^{-1}$.  Physically, the
reason is that the synchrotron spectrum decreases at higher
frequencies,  whereas the Black Body spectrum increases strongly.

\section{Conclusions}

We have provided detailed X-ray and radio studies of three radio
filaments near the Galactic center that have X-ray counterparts.  This
study suggests that two radio filaments with steep spectrum in radio
wavelengths are likely to be produced by inverse Compton scattering of
FIR radiation emitted from molecular clouds in the vicinity of the
filaments.  The inferred magnetic field strength from these
measurements indicate a field strength less than mG associated with
the filaments.  However, the largest uncertainty in these estimates is
the determination of dust temperature of the clouds.  Future high
resolution, sensitive far-IR and X-ray observations should be able to
investigate the role of inverse Compton scattering in the Galactic
center region which is known to have a mixture of thermal and
nonthermal gas.

% \appendix

% \section{}
% \label{}

% Bibliographic references with the natbib package:
% Parenthetical: \citep{Bai92} produces (Bailyn 1992).
% Textual: \citet{Bai95} produces Bailyn et al. (1995).
% An affix and part of a reference:
%   \citep[e.g.][Ch. 2]{Bar76}
%   produces (e.g. Barnes et al. 1976, Ch. 2).

\newpage
\begin{figure}
\caption{Contours of X-ray emission from
XMM J174540--2904.5
superimposed on a grayscale 2cm  continuum image of the SgrA-E, F and
G sources with a resolution of $2.27''\times1.23''$ (PA=-2$^0$). Contour
levels are set at (2, 3, 4, 6, 8, 12, 16, 20)$\times$ 1.5
$10^{-10}$ Jy beam$^{-1}$. The rms noise in the radio and X-ray images 
are 0.12 mJy beam$^{-1}$ and 1.8$\times10^{-11}$ Jy beam$^{-1}$, 
respectively.  
}
\end{figure}

\begin{figure}
\caption{Contours of 20cm   emission from SgrA-F
with a resolution of 2.17$''\times 1.05''$ (PA=9.7$^0$) are
superimposed on a grayscale X-ray image (in black).
The contour levels are -2, 2, 3, 4, 5, 6, 7, 8, 9, 10 mJy beam$^{-1}$
with rms noise 0.4 mJy beam$^{-1}$.
}
\end{figure}

\begin{figure}
\caption{(a) A full resolution X-ray image of Sgr A-E [top panel].
(b) A 2cm image of an identical region to (a) is shown on the
bottom panel. Both images have
a resolution of
$2.27''\times1.23''$
}
\end{figure}

\begin{figure}
\caption{The X-ray spectra of SgrA-E and F and their corresponding residuals.
Both spectra are featureless and are best fit by an absorbed
power law.}
\end{figure}

\begin{figure}
\caption{Contours of velocity integrated CS (2-1) 
line emission between 0 and
35 \kms superimposed  on a 6cm radio continuum image in black.  The
cross indicates the region corresponding to the spectrum in Fig.\ 6b. The 
molecular line data does not contain the single-dish spacings.}
\end{figure}

\begin{figure}
\caption{(a) A grayscale distribution of position-velocity diagram
of CS (2-1) line emission is shown on the top panel. (b)  A velocity
profile of peak CS (2-1) emission from the cross drawn on
on Figure 5 is shown on the bottom  panel. Both figures include 
single-dish spacings. }
\end{figure}

\begin{figure}
\caption{Contours of 1.2mm emission from the northern peak  of the
20 \kms molecular cloud M--0.13--0.08 set at (6, 6.5, 7,.. 10, 11, ..18, 20,
22, 24, 28, 32)$\times$0.16 Jy (Zylka et al. 1998) with the resolution of
10.6$''$. 
The grayscale 3.6cm  image  (in black)  shows 
 SgrA-E, F and G sources with the resolution of
$2.4''\times6.6''$ (PA=-0.9$^0$)}
\end{figure}

\begin{figure}
\caption{(a) (top panel) An adaptively smooth X-ray image of the filament with
the corresponding contours.
(bottom panel) Contours of X-ray emission superimposed on
a continuum image of
the brightest
portion of G359.54+0.18 at 6cm in J2000 coordinates. (b) The \emph{Chandra} X-ray spectrum of 
the ripple filament with an absorbed powerlaw fit.
}
\end{figure}

\begin{figure}
\caption{((a) top panel) The spectral index distribution between 
20 and 6cm
 whereas the bottom panel (b) shows the 
spectral index between 6 and 3.6cm. 
Both figures have 
selected the range of {\it uv} visibilities between 1 and 30 k$\lambda$,  
a final spatial resolution of 9.5$\Box''$. 
The cutoff values 
to determine the spectral index distribution in (a) are  2.2  and 0.41 mJy 
beam$^{-1}$ at 20 and 6cm, respectively. Similarly, for (b) the 
cutoff values are 0.41 and 0.24 mJy beam$^{-1}$  at 6 and 3.6cm. 
 These cut-off values are three times the rms noise of each 
of the images used to construct spectral index distribution.
}
\end{figure}

%\includegraphics[scale=0.42, angle=-90]{f2_spetra.ps}
%\end{center}

\end{document}